\documentclass[12pt, a4paper]{article}
\usepackage[T1,T2A]{fontenc}
\usepackage[utf8]{inputenc}

\usepackage{indentfirst}

 \usepackage{graphicx}
\usepackage{epsfig}

\begin{document}

\begin{center}
{\bfseries Optical properties of YAG:Ce and GGG:Ce scintillation crystals irradiated with a high fluence proton beam}
\vskip 5mm

V.~Baranov$^a$, Yu.I.~Davydov$^{a,}$\footnote{\small E-mail: davydov@jinr.ru},
M.~Mkrtchian$^b$, I.I.~Vasilyev$^a$

\vskip 5mm

{\small {\it $^a$ Joint Institute for Nuclear Research, Dubna, Russia}}\\
{\small {\it $^b$ A.Alikhanyan National Science Laboratory, Yerevan, Armenia}} \\
\end{center}

\vskip 5mm

\centerline{\bf Abstract}

\noindent

In this paper, we report on the study of the optical properties of YAG:Ce and GGG:Ce garnet crystals after irradiation in a 660~MeV proton beam with a fluence up to  8.9$\times$10$^{14}$ protons/cm$^2$. We found that the transparency of both crystals fell by no more than 7\% in the region of their own luminescence. The light yield of a YAG:Ce sample, measured one year after irradiation, dropped by about 35\% .

\vskip 10mm

\section{Introduction}
\label{intro}

Particle physics moves toward higher energies and beam intensities.
This will require the search for new materials and the development of detectors that provide high reliability and long-term stability in harsh radiation conditions.

Inorganic garnet crystals have recently been increasingly used  in tomography, medical applications, in particle physics and nuclear physics. This is due to their properties.  The structure of garnet crystal allows to change scintillation properties by doping the crystals with different ions~\cite{Kamada}.

Yttrium aluminum garnet crystals YAG:Ce (Y$_{3}$Al$_{5}$O$_{12}$:Ce) doped with Ce are candidates for use in future experiments in particle physics~\cite{Lucchini}.
Gadolinium gallium garnet crystals GGG:Ce (Gd$_{3}$Ga$_{5}$O$_{12}$:Ce)are widely used in the tomography, electronics~\cite{Kaminskii}, biology and medicine~\cite{Kaminska}.  

In this paper, we report on the results of our study of the optical properties of YAG:Ce and GGG:Ce garnet crystals before and after irradiation with a 660~MeV proton beam.

\section{Crystals}
\label{crystals}

Samples of yttrium aluminum garnet crystal YAG:Ce and gadolinium-gallium garnet crystal GGG:Ce doped with Ce were studied. The Ce content is 0.2at\% in YAG:Ce, and 0.3at\% in GGG:Ce. GGG:Ce crystal has a high density 7.08~g/cm$^{3}$. YAG:Ce crystal has a density 4.6~g/cm$^{3}$. Both crystals have luminescence in the 510-650~nm region with a peak of about 540~nm.
Crystal samples were cut from ingots grown by Czochralski method at the Institute for Physical Research~\cite{Ashtarak}, Ashtarak, Armenia. Crystal samples of both types had a size of 9$\times$9$\times$1~mm$^3$. All surfaces were optically polished.  The optical parameters of crystals were studied before irradiation and after irradiation with a proton beam.

\section{Proton beam}
\label{beam}

Crystal samples were irradiated in a 660~MeV proton beam at the phasotron of the Dzhelepov Laboratory of Nuclear Problems, JINR. The beam diameter was about 1.5~cm (FWHM). Samples were placed in the beam center. The total proton fluence through the samples was determined by measuring the induced activity in pure aluminum foil placed in the beam. The aluminum foil had a size of 30$\times$33 mm$^2$ and was cut into 110 pieces of 3$\times$3 mm$^2$ each, forming a 10$\times$11 matrix.

\begin{figure}[h]
\centering
\includegraphics[width=0.6\textwidth]{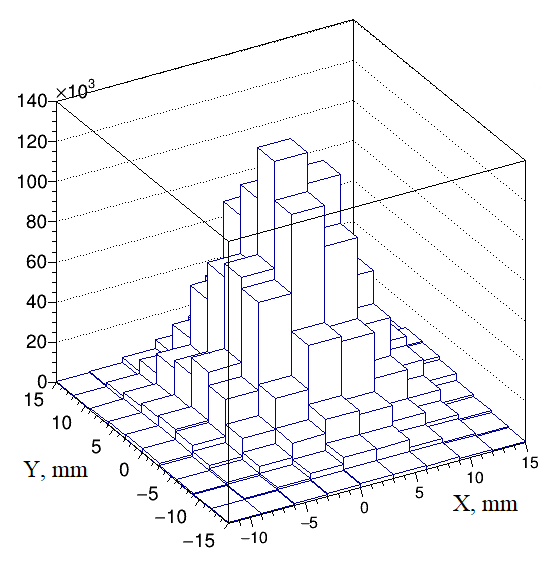}
\caption{Proton beam profile.}
\label{beam_profile}
\end{figure}

When protons pass through the aluminum foil, $^{24}$Na isotopes with a half-life of about 15 hours are produced in it. From a known cross section of the interaction of protons with aluminum nuclei and the measured induced activity of the foil, the fluence of protons passing through the foil is determined with a high accuracy. In our case, we used a different approach. The relative induced activity of each 3$\times$3 mm$^2$ aluminum piece was measured, the beam profile was constructed from these data, which was fitted with a three-dimensional Gaussian distribution. The beam profile
is shown in Figure~\ref{beam_profile}. Then the absolute values for each 3$\times$3 mm$^2$ aluminum piece were determined from the accelerator beam current during the irradiation of the samples.  As a result of the fit, it was found that 31\% of the total fluence of the proton beam passes through the central part of 9$\times$9 mm$^2$, where the samples were placed during the irradiation.

Two exposures of samples have been performed. During the first irradiation, 2.3$\times$10$^{13}$ protons or 2.8$\times$10$^{13}$ protons/cm$^2$ were passed through the samples. The second exposure added 69.7$\times$10$^{13}$ protons, giving about 7.2$\times$10$^{14}$ protons on sample or 8.9$\times$10$^{14}$ protons/cm$^2$ in total.

\section{Results and discussion}
\label{results}

\begin{figure}[h]
\centering
\includegraphics[width=0.45\textwidth]{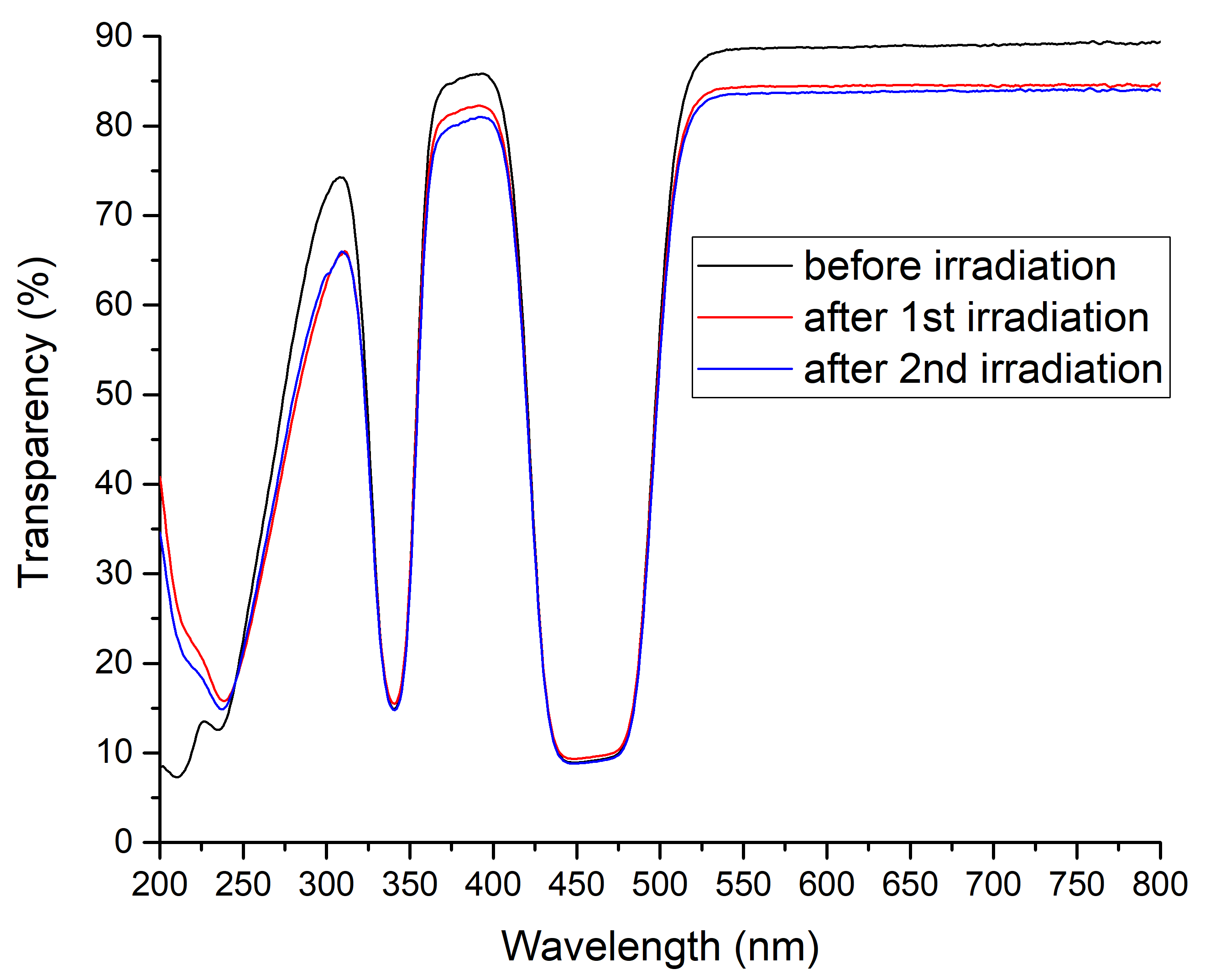}
\includegraphics[width=0.45\textwidth]{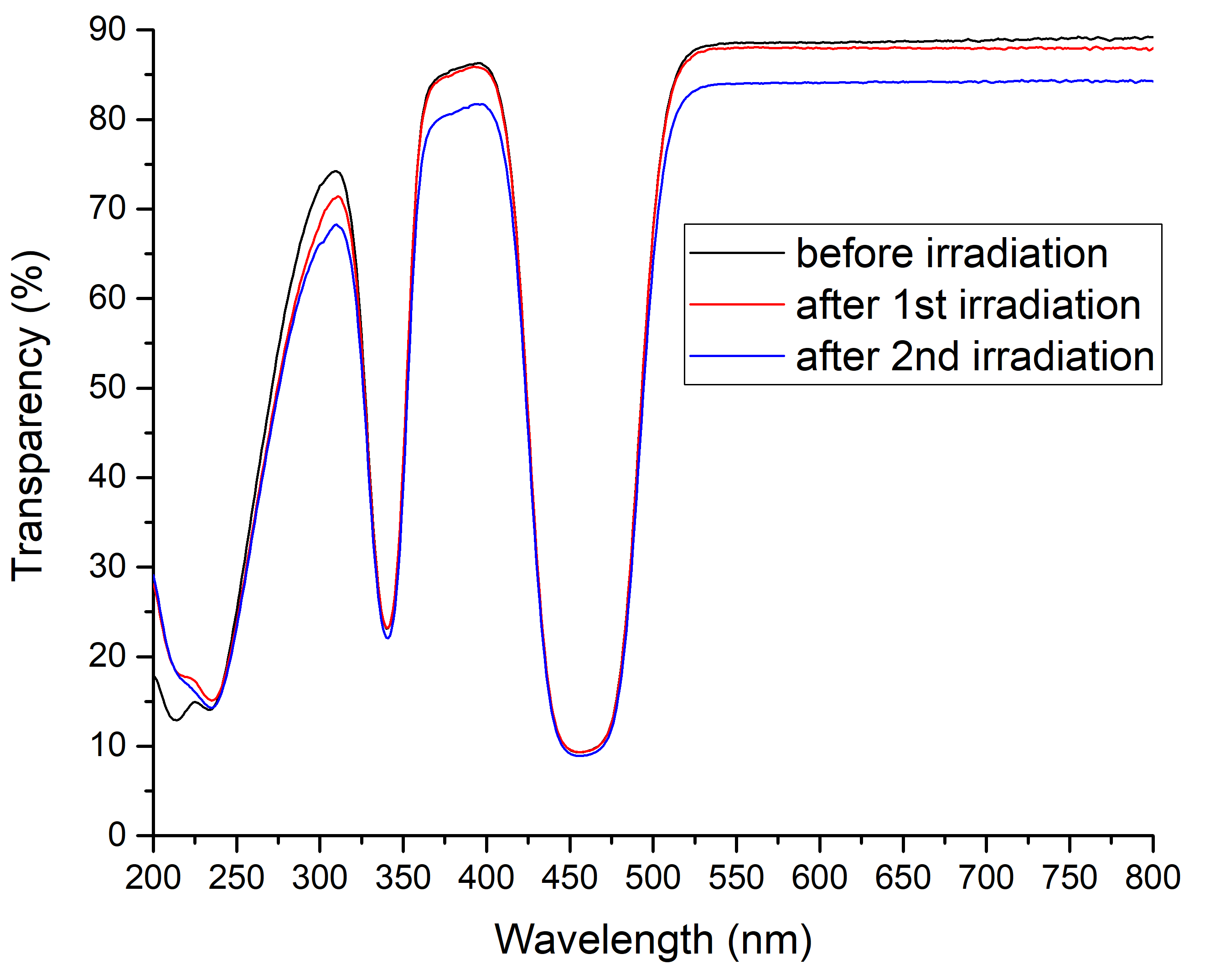}
\caption{Transmittance of the YAG:Ce crystal (on the left) and GGG:Ce crystal (on the right) before and after irradiation. }
\label{YAG_transmit}
\end{figure}

The transmission of the samples was measured using a spectrophotometer Shimadzu SolidSpec-3700 DUV. Light transmission of YAG:Ce and GGG:Ce samples before and after irradiation are shown in Fig.\ref{YAG_transmit}.
The measurement results demonstrated that after the second irradiation, the transparency of both crystal samples fell by no more than 7\% in the region of their own luminescence (approximately 510-650 nm).  However, there are differences in the transparency behavior of the two crystals  after first and second exposures. The transparency of the GGG:Ce sample fеll mainly after the second exposure, while a significant decrease in the transparency of the YAG:Ce sample occurred immediately after the first exposure.

\begin{figure}[h]
\centering
\includegraphics[width=0.6\textwidth]{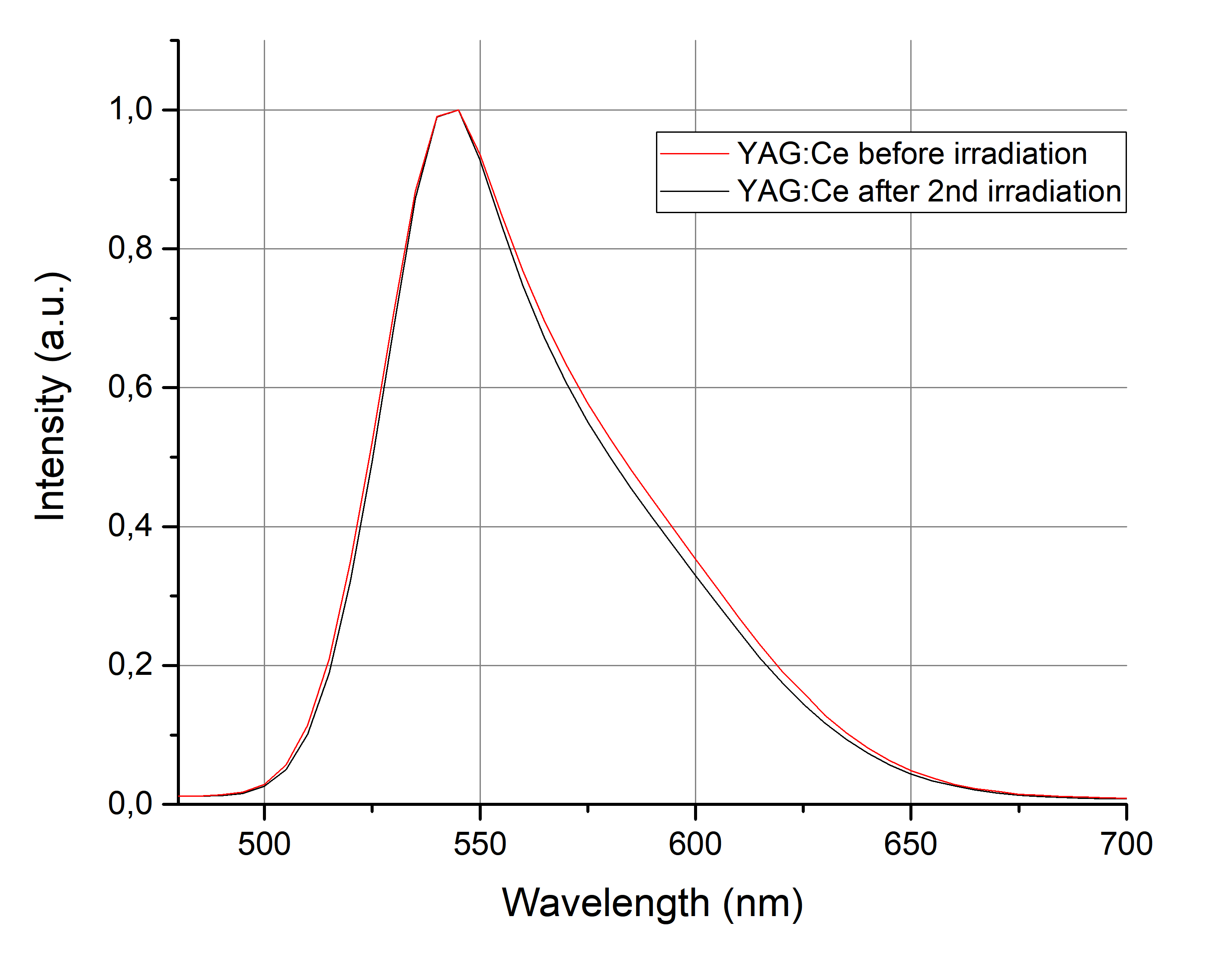}
\caption{YAG:Ce luminescence spectra before and after irradiation.}
\label{YAG_luminescence}
\end{figure}

Luminescence spectra of the samples before and after irradiation were measured with a monochromator SF-46. Both samples have a luminescence peak around of 540~nm. As an example, the normalized luminescence spectra of a YAG:Ce crystal before and after the second irradiation are shown on the Figure~\ref{YAG_luminescence}. The luminescence of the irradiated sample was measured one year after irradiation. One can see that the shape of the luminescence spectrum of the crystal did not change when 7.2$\times$10$^{14}$ protons pass through the sample.

The light output of the samples before and after second irradiation was estimated by the response of the crystals to the beta source of $^{90}$Sr. The decay electrons passed through the crystal under study and fell into a trigger counter made of a larger YAG:Ce crystal. The registration of energy deposited in the trigger counter allowed us to select electrons intersecting the crystals under study with an energy of about 1.5–2.28~MeV (the beta decay spectrum continues up to 2.28 MeV). The change in the light output of investigated crystals was determined by the peak position in the spectrum of the electrons from the $^{90}$Sr beta decay intersecting crystals.

\begin{figure}[ht]
\centering
\includegraphics[width=0.6\textwidth]{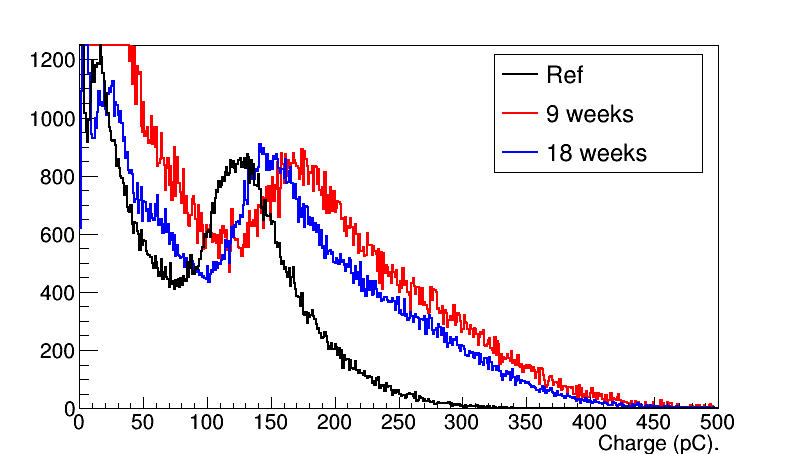}
\caption{Spectra from non-irradiated (black line) and irradiated (red and blue lines) samples of the YAG:Ce crystal due to $^{90}$Sr beta source. Spectra from the irradiated sample were taken 9 (red line) and 18 (blue line) weeks after the irradiation.}
\label{YAG_beta}
\end{figure}

\begin{figure}[ht]
\centering
\includegraphics[width=0.6\textwidth]{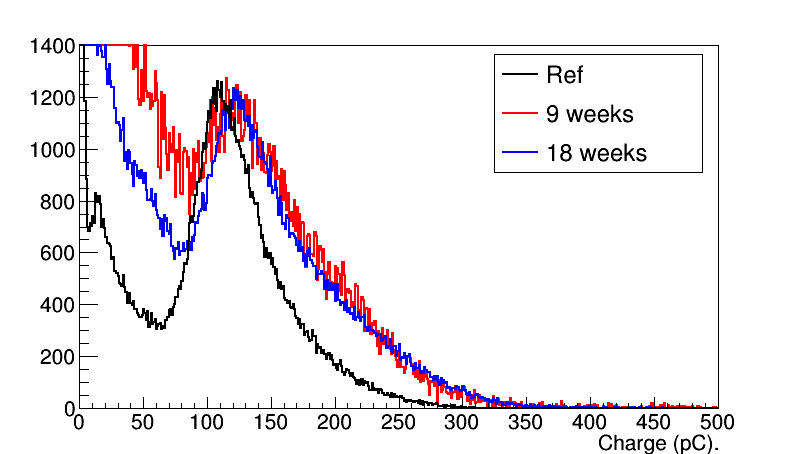}
\caption{Spectra from non-irradiated (black line) and irradiated (red and blue lines) samples of the GGG:Ce crystal due to $^{90}$Sr beta source. Spectra from the irradiated sample were taken 9 (red line) and 18 (blue line) weeks after the irradiation.}
\label{GGG_beta}
\end{figure}

Protons crossing the crystal samples produce radioisotopes in the crystal due to nuclear reactions~\cite{Auffray}. The high residual activity of the crystals after irradiation greatly complicates the measurement of the spectrum of beta decay electrons. First measurements of the irradiated samples were made 9 weeks after the exposure.

Earlier, we developed and tested a method for extracting the contribution from an external source against the background. When subtracting two spectra, from intrinsic radioactivity and from external source plus intrinsic radioactivity, taking into account the fraction of the external source and the intrinsic radioactivity, the contribution from the external source is reliably distinguished even with its relatively small fraction in the total spectrum~\cite{Afanaciev}.

Figure~\ref{YAG_beta} shows the spectra from the non-irradiated YAG:Ce crystal (black line) and after the second irradiation (red and blue lines) due to $^{90}$Sr beta source. The beta source contribution to the spectrum of the irradiated sample was obtained by subtracting two spectra:  the residual activity spectrum was subtracted from the total spectrum due to residual activity plus external irradiation by the $^{90}$Sr beta source.

After irradiation, both samples were found to show an increase in light output when irradiated with $^{90}$Sr beta source. The first measurement of YAG:Ce sample, taken 9 weeks after the irradiation, shows signals increase of almost 30\% (red line) compared with the non-irradiated samples. After 18 weeks, the output signals from the irradiated sample (blue line) were still higher than that from the non-irradiated samples, as seen on the Figure~\ref{YAG_beta} for the YAG:Ce crystal. The increase in light output  from a GGG:Ce crystal after irradiation was much smaller than for YAG:Ce sample, but is still visible. Spectra from the GGG:Ce crystal are presented in Fig.~\ref{GGG_beta}. The irradiated sample was measured 9 and 18 weeks after exposure as well. We observed such an effect earlier on some crystals. Apparently, this is due to radioisotopes and excited states arising in samples as a result of irradiation with protons.

\begin{figure}[h]
\centering
\includegraphics[width=0.45\textwidth]{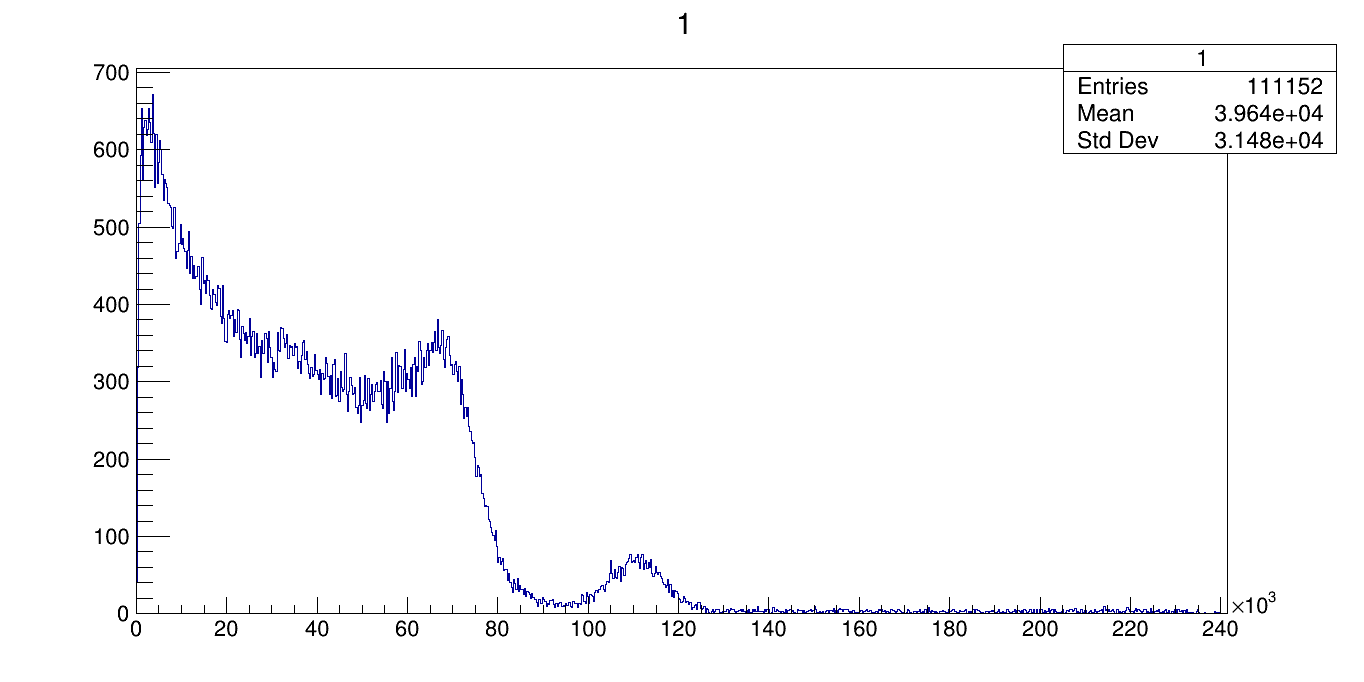}
\includegraphics[width=0.45\textwidth]{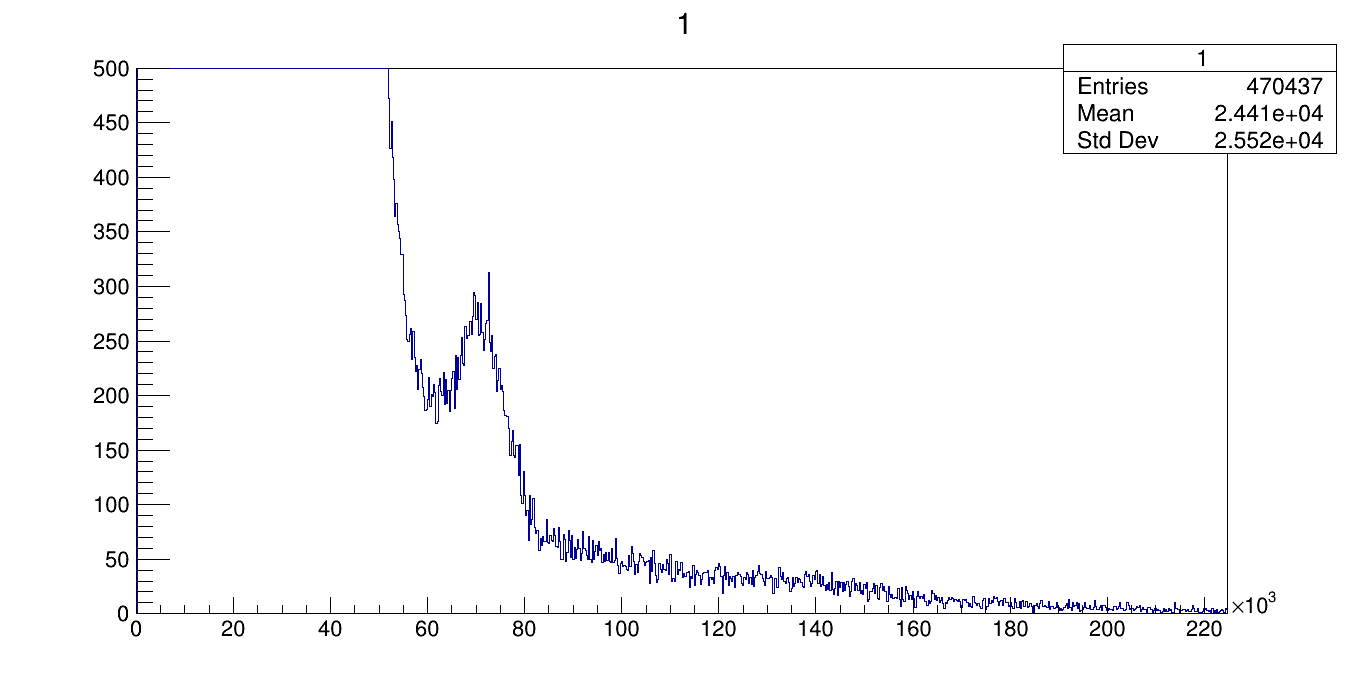}
\caption{Spectra from non-irradiated (left) and irradiated (right) samples of the YAG:Ce crystal due to $^{22}$Na gamma source. }
\label{YAG_gamma}
\end{figure}

Data taken from both YAG:Ce and GGG:Ce samples several months after irradiation show that eventually light output from irradiated samples became lower compared to non-irradiated samples. Figure~\ref{YAG_gamma} shows the response of non-irradiated and irradiated YAG:Ce samples to the $^{22}$Na gamma source. The spectrum from the irradiated sample was taken about a year after irradiation. The full absorption peak shows that the irradiated YAG:Ce sample lost about 30\% of the light output compared with the non-irradiated sample.

\section{Conclusion}

The optical properties of YAG:Ce and GGG:Ce garnet crystals were studied before and after irradiation in a 660~MeV proton beam. Approximately 7.2$\times$10$^{14}$ protons per sample or 8.9$\times$10$^{14}$ protons/cm$^2$ in total passed through the samples.
The transparency of both crystals fell by no more than 7\% in the region of their own luminescence.
Both samples received high induced activity due to proton-induced nuclear reactions. As a result, both irradiated crystals had a higher light output for several months after irradiation than non-irradiated crystals. Ultimately, the irradiated samples cooled down and after a year YAG:Ce sample shows a light output of about 35\% less  compared to  non-irradiated sample.

\section*{Acknowledgment}

The authors are grateful to Dr.~G.V.~Mitsyn and Dr.~A.G.~Molokanov for help in irradiating the samples.

The reported study was funded by the Russian Fund for Basic Research, project number 18-52-05021.

\end{document}